\documentclass[aps,prb,superscriptaddress,twocolumn,showpacs,amsmath,amssymb,reprint,longbibliography]{revtex4-2}

\usepackage{graphicx}
\usepackage{dcolumn}
\usepackage{bm}
\usepackage{lipsum}
\usepackage{chemformula}
\usepackage[T1]{fontenc}
\usepackage[colorlinks=true, allcolors=blue]{hyperref}
\usepackage{amsfonts}
\usepackage{algorithm2e}
\usepackage[utf8]{inputenc}
\usepackage[T1]{fontenc}
\usepackage{mathptmx}
\usepackage{ulem}

\begin{document}


\title{Discovery of an ultrastable antiferromagnetic two-dimensional CrF$_{3}$ phase with anisotropic quasi-one-dimensional mechanical, electronic, and thermal properties}

\author{Xin Chen}
\affiliation{Thermal Science Research Center, Shandong Institute of Advanced Technology, Jinan 250100, Shandong Province, People’s Republic of China}

\author{Fengyi Zhou}
\affiliation{ 
Faculty of Applied Sciences, Macao Polytechnic University, Macao SAR, 999078, People’s Republic of China
}%
\author{Yan Suo}
\affiliation{ 
Faculty of Applied Sciences, Macao Polytechnic University, Macao SAR, 999078, People’s Republic of China
}%
\author{Cheng Shao}
\affiliation{Thermal Science Research Center, Shandong Institute of Advanced Technology, Jinan 250100, Shandong Province, People’s Republic of China}

\author{Xu Cheng}
\email{xchengab@sdu.edu.cn}
\affiliation{
Institute for Advanced Technology, Shandong University, Jinan 250061, People's Republic of China
}

\author{Duo Wang}
\email{duo.wang@mpu.edu.mo}
\affiliation{ 
Faculty of Applied Sciences, Macao Polytechnic University, Macao SAR, 999078, People’s Republic of China
}

\author{Biplab Sanyal}
\email{biplab.sanyal@physics.uu.se}
\affiliation{Department of Physics and Astronomy, Uppsala University, Box 516,
751\,20 Uppsala, Sweden}
\date{\today}

\begin{abstract}
We report the discovery of an ultra-stable antiferromagnetic two-dimensional (2D) CrF$_3$ phase that is energetically more favorable than the traditionally assumed hexagonal structure. Using first-principles calculations combined with evolutionary structure searches, we identify a new low-energy rectangular configuration of CrF$_3$, which exhibits remarkable anisotropic properties. Mechanically, this phase features zero in-plane Poisson’s ratio, a rare negative out-of-plane Poisson’s ratio, and quasi-one-dimensional (quasi-1D) mechanical behavior, characterized by minimal coupling between orthogonal directions. Electronically, CrF$_3$ demonstrates quasi-1D transport properties with two independent conduction bands near the Fermi level, which are selectively tunable by applying uniaxial strain. The calculated bandgap is 3.05 eV, and it can be continuously modulated under strain, providing a controllable means to engineer its electronic properties. The material also displays out-of-plane antiferromagnetic ordering with a magnetic anisotropy energy of 0.098 meV/Cr atom and an estimated Néel temperature of 24 K. In addition, we investigate the thermal conductivity of monolayer rectangular CrF$_3$ (r-CrF$_{3}$), revealing significant anisotropy in its heat transport properties. The thermal conductivity along the y-axis is approximately 60.5 W/mK at 300 K, much higher than along the x-axis, where it is only 13.2 W/mK. The thermal anisotropic factor is computed to be 4.58, highlighting the material’s strong directional heat transport capabilities. This factor surpasses that of other 2D materials such as black phosphorene, WTe$_{2}$, and arsenene, indicating the potential of r-CrF$_3$ for advanced directional heat management applications. This makes the rectangular CrF$_3$ phase a promising candidate for applications in spintronics, strain-engineered nanoelectronics, mechanical metamaterials, and thermal management technologies.
\end{abstract}

\maketitle


\section{Introduction}
Since the realization of graphene \cite{Novoselov666}, two-dimensional (2D) materials have garnered extensive research interest due to their novel physical properties and potential applications in nanoelectronics and spintronics \cite{Zhu2015,doi:10.1021/nn501226z,doi:10.1002/anie.201411246,PhysRevB.87.165415,doi:10.1021/nl903868w}. Beyond graphene, a plethora of other 2D materials have been explored, \cite{Huang2018,doi:10.1021/cm504242t,Lado_2017,doi:10.1021/ja508154e,doi:10.1021/jacs.7b06296}, with recent attention on 2D magnetic materials \cite{doi:10.1126/science.aav4450,Gibertini2019,Mak2019,10.1063/5.0147450}. Among these, chromium triiodide (CrI$_{3}$) stands out as one of the earliest identified 2D ferromagnetic materials exhibiting out-of-plane magnetization, attracting significant interest due to its rich physical phenomena \cite{Huang2017,Kashin_2020,PhysRevB.99.104432}. For instance, the magnetic properties of CrI$_{3}$ vary with the number of layers, offering possibilities for designing novel spintronic devices \cite{Huang2017,doi:10.1021/acs.jpcc.1c04311}. Furthermore, heterostructures formed by stacking CrI$_{3}$ with other 2D materials such as graphene and transition metal dichalcogenides (TMDCs) have unveiled intriguing physical behaviors and potential applications. 

Within the family of transition metal halides, chromium trifluoride (CrF$_{3}$), along with CrCl$_{3}$ and CrBr$_{3}$, has also been recognized as a potential two-dimensional ferromagnetic semiconductor material \cite{PhysRevB.102.115162,PhysRevB.99.104432,doi:10.1126/science.aah6015,PhysRevLett.106.167206,Acharya2022,C5CP04835D,Wines2023,C5TC02840J,https://doi.org/10.1002/smll.202300333,TOMAR2019165384}. Traditionally, CrF$_{3}$ has been assumed to adopt a hexagonal ferromagnetic structure similar to that of CrI$_{3}$, and numerous theoretical studies have been conducted based on this assumption \cite{zheng2024spinlatticecouplinginducedrich,C5TC02840J,TOMAR2019165384,https://doi.org/10.1002/smll.202300333,Acharya2022,liu2022group,Wines2023}. These studies have investigated various properties of hexagonal CrF$_{3}$, including its electronic structure, magnetic ordering, and optical characteristics. Compared with Cl, Br, and I compounds, CrF$_3$ have many special properties, such as it could host the most localized Frenkel-like excitons with the largest binding energies \cite{Acharya2022}. However, despite extensive theoretical interest, experimental realization of 2D CrF$_{3}$ with the hexagonal ferromagnetic structure has remained elusive, suggesting that this assumed structure may not represent the most stable configuration for CrF$_{3}$ in two dimensions.

In the realm of first-principles simulations, obtaining reliable structural information is a fundamental prerequisite for subsequent investigations of material properties. However, even for many synthesized materials, acquiring comprehensive structural details remains challenging due to experimental limitations. For materials yet to be synthesized, researchers often rely on empirical conjectures. For example, the assumption that CrF$_{3}$ shares a hexagonal structure similar to its homologous compound CrI$_{3}$ is an experiential judgment based on analogy. In recent years, several algorithms have been proposed to address this fundamental challenge, including simulated annealing, genetic algorithms, data mining, and metadynamics \cite{Pannetier1990,doi:10.1002/anie.199612861,PhysRevLett.90.075503,C2CE06642D,PhysRevLett.91.135503}. Among these methods, evolutionary algorithms—where a population of candidate structures evolves through iterations of random variation and selection have proven to be particularly effective \cite{LYAKHOV20131172,doi:10.1063/1.2210932}. A comprehensive discussion of the operational principles behind this algorithm and its efficacy in structure prediction can be found in Ref. \cite{doi:10.1021/ar1001318}. The application of evolutionary algorithms has facilitated the prediction of numerous new materials characterized by excellent stability and intriguing properties \cite{Zhang1502,Oganov2019,PhysRevLett.112.085502,PhysRevB.87.195317,PhysRevLett.110.136403,Zhang2017}.

Quasi-one-dimensional (quasi-1D) materials have emerged as technologically appealing platforms for studying quantum phenomena that are challenging to realize in strictly one-dimensional (1D) systems due to instability, scalability issues, and ensemble inhomogeneities. While experimental realizations of true 1D systems—such as ultracold quantum gases \cite{Paredes2004}, atomic chains \cite{Blumenstein2011}, and carbon nanotubes \cite{doi:10.1126/science.275.5297.187} —have enabled exploration of correlated phenomena like Tomonaga–Luttinger liquids \cite{Blumenstein2011}, their practical limitations hinder broader applications. Quasi-1D materials offer an alternative by exhibiting strong anisotropy in their atomic arrangements and electronic interactions along different crystallographic directions. Notable examples include MoS$_{2}$ \cite{doi:10.1126/science.1059011}, III–V-based nanowires \cite{PhysRevB.104.235307}. These materials are particularly intriguing when they possess both electronic and magnetic properties, as they can host exotic quantum phenomena and functionalities. For instance, the recent observation of long-range magnetic order in van der Waals magnets like TiOCl and CrSBr has opened prospects for materials exhibiting correlated excitations, including electron–electron interactions, spin dynamics, and electron–phonon coupling \cite{PhysRevLett.102.056406, Klein2023}. Additionally, materials such as NbSi$_{x}$Te$_{2}$ \cite {Zhang2022} and rectangular C$_{4}$N monolayer \cite{adfm.202415606} demonstrate that by tuning structural parameters, one can manipulate the dimensionality of Dirac fermions, thereby providing platforms to study quasi-1D Dirac physics relevant for high-speed electronic applications.

In this work, we employ first-principles calculations combined with an evolutionary structure search to investigate the stable phases of two-dimensional CrF$_3$. Our results reveal a novel rectangular phase (r-CrF$_3$) that is more stable than the widely assumed hexagonal phase (h-CrF$_3$), fundamentally challenging the conventional belief that CrF$_3$ adopts a hexagonal structure similar to CrI$_3$ and its halide counterparts (collectively referred to as h-CrX$_{3}$ in the following, where X = F, Cl, Br, I). This new phase exhibits remarkable quasi-1D behavior, characterized by nearly zero in-plane Poisson’s ratio and minimal coupling between orthogonal directions, where strain along one axis induces negligible deformation in the perpendicular direction. The rectangular structure also displays out-of-plane antiferromagnetic ordering with a magnetic anisotropy energy of 0.098 meV/Cr atom and an estimated Néel temperature of 24 K, indicating its potential for low-temperature spintronic applications. Moreover, under tensile strain applied in the x-direction or [120]] direction, r-CrF$_3$ demonstrates a rare negative Poisson’s ratio (NPR), making it a promising candidate for mechanical metamaterials. Additionally, r-CrF$_3$ exhibits highly anisotropic thermal conductivity, with significantly higher heat transport along the y-direction (60.5 W/mK at 300 K) compared to the x-direction (13.2 W/mK), making it suitable for directional heat management. The electronic band structure of r-CrF$_3$ features two independent conduction bands near the Fermi level: the lower-energy band is highly sensitive to strain along the x-direction, while the higher-energy band is primarily tuned by strain in the y-direction, enabling precise and strain-tunable electronic properties. Combined with its robust out-of-plane antiferromagnetic order and significant magnetic anisotropy energy, these findings not only revise the understanding of CrF$_3$ but also highlight r-CrF$_3$ as a multifunctional material with potential applications in strain-engineered electronics, mechanical metamaterials, and spintronic devices.

\begin{figure}[hbtp]
\includegraphics[scale=1]{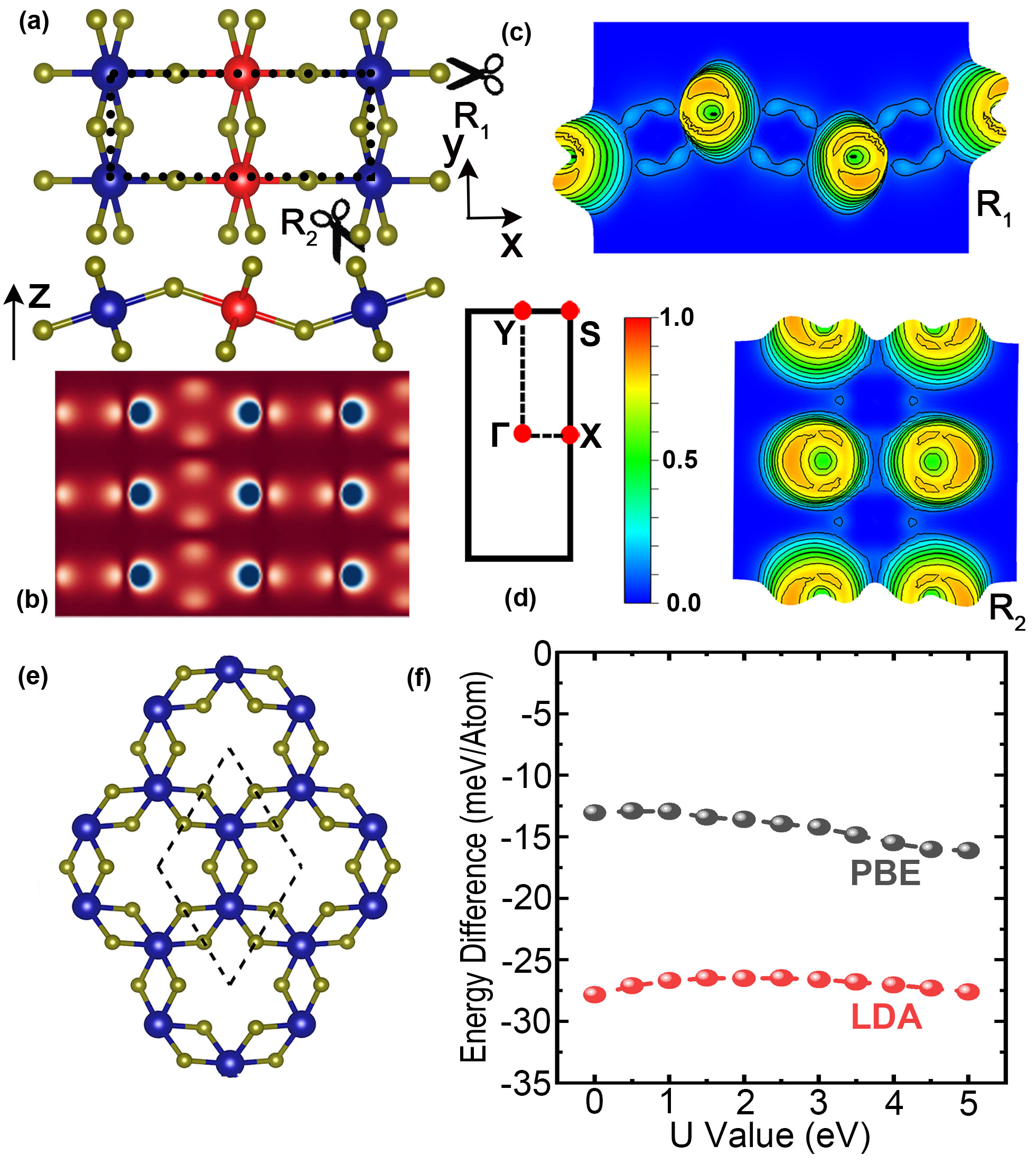} 
\caption{(a) The top and side views of the atomic structures of r-CrF$_{3}$ monolayer with its primitive cell marked by dashed square and the cell axes indicated by black arrows. Cr atoms with different magnetic moment are in blue and red respectively. (b) A typical simulated STM image of 2D r-CrF$_{3}$. (c) The contour maps of ELFs of r-CrF$_{3}$, which are sliced on F-Cr-F planes marked with scissors R$_1$ and R$_{2}$ in (a). (d) The dashed black square is the first BZ of r-CrF$_{3}$ monolayers, with the high symmetric points indicated.  (e) The structure of h-CrF$_{3}$. (f) The energy difference between r-CrF$_{3}$ and h-CrF$_{3}$ as a function of Hubbard U parameters. The red symbols and lines denote the LDA results, while the gray symbols and lines represent the PBE results.
}
\label{figure1}
\end{figure}

\section{Computational details}
We performed the global structure search with the evolutionary algorithm based code USPEX \cite{doi:10.1063/1.2210932,GLASS2006713,PhysRevB.87.195317,doi:10.1021/ar1001318,LYAKHOV20131172} to obtain low energy stable structures. In the structure search, different magnetic configurations have been considered. All the first-principles calculations are performed by using the density functional code VASP\cite{002230939500355X,PhysRevB.54.11169}. The exchange-correlation-functional was approximated by local spin density approximation considering Hubbard U formalism for correlated Cr-d orbitals (LSDA+U), and the U value was set as 3.0 eV. For the calculations of 2D materials, the out-of-plane interaction is avoided by taking a vacuum of more than 20 \AA. The plane-wave energy cutoff was set to a minimum of 520 eV, and the energy convergence criterion was set to 10$^{-6}$ eV. Structural optimization was performed until the atomic forces were reduced to less than 0.001 eV/\AA. The Brillouin zone (BZ) was sampled using a Monkhorst-Pack grid with a density of no less than $2 \pi \times 0.033$ \AA$^{-1}$. To assess the dynamical stability of the structures, phonon spectra were calculated using the PHONOPY code \cite{phonopy}. Ab initio molecular dynamics (AIMD) simulations were conducted at a temperature of 1200 K for up to 10 ps. The interatomic exchange parameters were calculated using the full-potential linear muffin-tin orbital (FP-LMTO) code RSPt \cite{wills2010full}, after which an effective spin Hamiltonian was constructed as follows:
\begin{equation} \hat{H}=-\sum_{i \neq j} J_{ij} \Vec{S_{i}} \cdot \Vec{S_{j}} - \sum_i K_i (S_i^z)^2, \end{equation}
where $\Vec{S_{i}}$ is the magnetic moment along the magnetization direction at site $i$, $J_{ij}$ represents the exchange interaction between two sites, and $K_i$ denotes the strength of single ion magnetocrystalline anisotropy energy (MAE) at site $i$. 
The magnetic phase transition temperature was then obtained using classical Monte Carlo simulations, as it is implemented in the UppASD package~\cite{skubic2008method}. 
With a lattice size of $50\times 120\times 1$ (containing 12000 Cr atoms) and periodic boundary conditions applied in two dimensions, an annealing process was simulated, starting at 250 K and gradually cooling to 0 K, with 25,000 Monte Carlo steps performed at each temperature.
For comparison, we also use the calculated exchange parameters to evaluate the ordering temperature based on the mean-field approximation (MFA) ~\cite{kvashnin2015exchange}: $T_c=\frac{2J_0}{3k_B}$, where $J_0=\sum_j J_{0j}$ corresponds to the sum of the exchange interactions parameters. Post-processing of the calculations was performed using VASPKIT \cite{wang2019vaspkit} and VESTA \cite{Mommako5060}.

The lattice thermal conductivity of the 2D material is computed using this formula \cite{C8NR01649F}:
\begin{equation}
\kappa_{\alpha \beta} = \frac{1}{SH} \sum_{\lambda} C_{\lambda} v_{\lambda \alpha} v_{\lambda \beta} \tau_{\lambda},
\end{equation}
in which \(S\) is the surface area, \(\lambda\) denotes the phonon modes, \(C_{\lambda}\) is the heat capacity of mode \(\lambda\), \(v_{\lambda \alpha}\) (\(v_{\lambda \beta}\)) is the group velocity in the \(\alpha\) (\(\beta\))-direction, and \(\tau_{\lambda}\) is the relaxation time. \(H\) is calculated as \(H = h_{CrF_{3}} + 2 r_{F}\), where \(h_{CrF_{3}}\) is thickness of the slab, and \(r_{F}\) is the van der Waals radius of the \(F\) atom \cite{doi:10.1021/j100785a001}. The results are obtained by solving the phonon Boltzmann transport equation (BTE) iteratively using the ShengBTE code \cite{ShengBTE_2014,PhysRevB.85.195436}. The second- and third-order force constants are calculated using density functional perturbation theory (DFPT) and the Lagrange multiplier method \cite{PhysRevB.86.174307}, respectively. For the force-constant calculations, a \(3 \times 3 \times 1\) supercell are used, including interactions up to the fifth nearest-neighbor atoms. In the BTE calculations, we employed a \(\Gamma\)-centered \(16 \times 30 \times 1\) q-mesh for convergence.

\section{Results and discussion}
After a systematic structure search, the CrF$_{3}$ monolayer with a rectangular structure [space group Pmma (No. 51)], as shown in Fig. \ref{figure1} (a) and referred to as r-CrF$_{3}$, is found to be energetically favored. There are two Cr atoms with antiparallel magnetic moment in the unitcell, and all of the Cr atoms are connected with F atoms and braided into the whole 2D lattice. As shown in the side view, neighbor Cr atoms are connected by single F atoms along the $x$-direction and by pairs of F atoms along the y-direction. Specifically, the single F atoms along the x-direction are not located in the same plane as the Cr atoms, resulting in a staggered configuration between the Cr and F atoms. Additionally, the F-F bonds between the pairs of F atoms in the y-direction are tilted, and their orientation is not parallel to the z-axis, leading to a non-collinear arrangement of the fluorine atoms. The lattice constants of r-CrF$_{3}$ are a$_{x}$ = 7.213 \AA\ and a${_y}$ = 2.939 \AA, while the monolayer thickness is h = 2.341 \AA. To aid in experimental comparisons, we have simulated the scanning tunneling microscope (STM) image of the r-CrF$_{3}$ monolayer, which is shown in Fig. \ref{figure1} (b). 

\begin{figure}[htbp]
\includegraphics[scale=1]{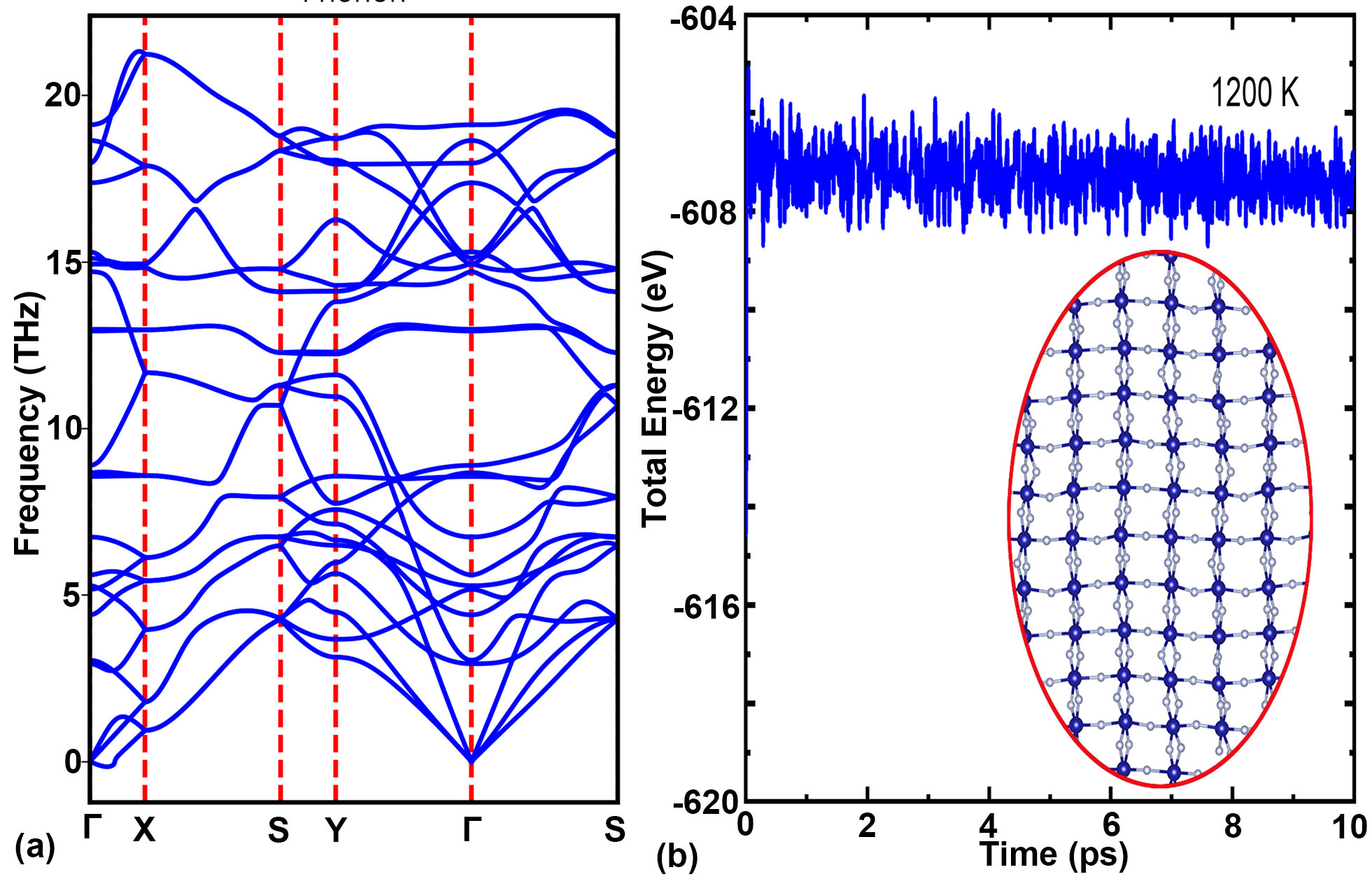}
\caption{(a) The phonon spectra along the high symmetry path in the first Brillouin zone. (b) The Total energy of r-CrF$_{3}$ as a function of time.}
\label{figure2}
\end{figure}

We calculated the electron localization function (ELF) for analyzing the chemical bond nature of r-CrF$_{3}$ monolayer, which are shown in Figure \ref{figure1} (c). According to the calculations, in r-CrF$_{3}$ monolayer, the electrons are mostly localized near F atoms, implying that F atoms gain electrons from Cr atoms, forming ionic bonds, which is also supported by the Bader charge analysis in Table \ref{table1}. Compared with h-CrF$_{3}$, r-CrF$_{3}$ have less transferred Bader charge, showing that the ionization is weaker in r-CrF$_{3}$ than in r-CrF$_{3}$. 

Notably, the entropy of r-CrF$_{3}$ is found to be 27 meV/atom lower than the reported h-CrF$_{3}$ as shown in Fig. \ref{figure1} (e), which shows P-31m space group symmetry and were assumed as the stable phase of CrF$_{3}$ in relevant works. We also validated the robustness of our findings by examining different Hubbard $U$ parameters and exchange-correlation functionals. As shown in Fig.~\ref{figure1}(f), under LDA+U, r-CrF$_{3}$ consistently exhibits a lower total energy than the hexagonal phase across all tested $U$ values, with an energy difference of 26.5--27.8\,meV/atom. Meanwhile, our additional calculations using PBE+U predict slightly larger lattice constants but confirm the same energetic ordering, yielding an energy difference of 12.9--13.1\,meV/atom. These results further reinforce that r-CrF$_{3}$ remains the more stable phase under various computational settings.


The dynamic stability of r-CrF$_{3}$ is confirmed by the density functional perturbation theory calculations and ab initio Born-Oppenheimer molecular dynamics (BOMD) simulations. In the DFPT calculations, we employed a $3 \times 3 \times 1$ supercell. The phonon spectra along the high symmetry path in the first BZ indicated in Fig. \ref{figure1} (d) and shown in Fig. \ref{figure2} (a). We note that there is a very small imaginary frequency pocket near the $\Gamma$ point in the out-of-plane acoustic branch, which is frequently reported for two-dimensional materials\cite{doi:10.1021/acs.nanolett.6b00070,PhysRevMaterials.3.074002,C7TC00789B}. Such tiny negative values are generally recognized as numerical artifacts resulting from the difficulty of converging long-wavelength bending modes in 2D systems, and do not necessarily indicate a real dynamical instability\cite{PhysRevB.94.165432}. To further verify the stability, we performed BOMD simulations on a $2 \times 6 \times 1$ supercells of r-CrF$_{3}$ at 1200 K until 10 ps using a 1 fs time step. As shown in Fig. \ref{figure2}(b), the total energies of r-CrF$_{3}$ monolayer remains stable, and there is no obvious structural change in the final geometrical framework.


To study the energetic stability and compare to other 2D materials, we compute the cohesive energy following the expression
\begin{equation}
\Delta E=\frac{ E_{CrF_{3}}-(n_{Cr} \times E_{Cr}+n_{F} \times E_{F}) }{n_{Cr}+n_{F}}.
\label{eq1}
\end{equation}
In this formula, E$_{CrF_{3}}$ is the total energy of r-/h-CrF$_{3}$, E$_{Cr}$, and E$_{F}$ are the energies of Cr and F atoms. The values of $\Delta E$ are listed in Table \ref{table1}. For comparison, the values of the synthesized germanene and stanene computed, as -3.26 eV/atom and -2.74 eV/atom, respectively \cite{doi:10.1021/ja513209c}. The low cohesive energy implies that the synthesis of r-CrF$_{3}$ is highly probable.

\begin{table}[htbp]
\begin{center}
\caption{The lattice parameters (a$_{x}$ and a$_{y}$, in \AA), slab thicknesses (h, in \AA), the Bader charge transfers from Cr to F atoms ($\Delta \rho_{Cr}$, in e/(Cr atom)) of CrF$_{3}$ structures, from Cr to F1 atoms ($\Delta \rho_{F1}$, in e/(F atom)), from Cr to F2 atoms ($\Delta \rho_{F2}$, in e/(F atom)), average Bader charge transfer from Cr to F aotms ($\Delta \rho_{\overline F}$, in e/(F atom)), and cohesive energies ($\Delta E$, in eV/atom) of r-CrF$_{3}$ and h-CrF$_{3}$ monolayers.}
\label{table1}
\tabcolsep2pt
\arrayrulewidth0.5pt
\begin{tabular}{|c|c|c|c|c|c|c|c|c|}
\hline
\hline Phase&a$_{x}$&a$_{y}$&h&$\Delta \rho_{Cr}$&$\Delta \rho_{F1}$&$\Delta \rho_{F2}$&$\Delta \rho_{\overline F}$&$\Delta E$\\
\hline r-CrF$_{3}$&7.215&2.939&2.342&1.723&0.590&0.567&0.574&-4.878\\
\hline h-CrF$_{3}$&5.072&-&2.024&1.786&-&-&0.595&-4.851\\
\hline
\hline
\end{tabular}
\end{center}
\end{table}

\begin{figure}[htbp]
\includegraphics[scale=0.35]{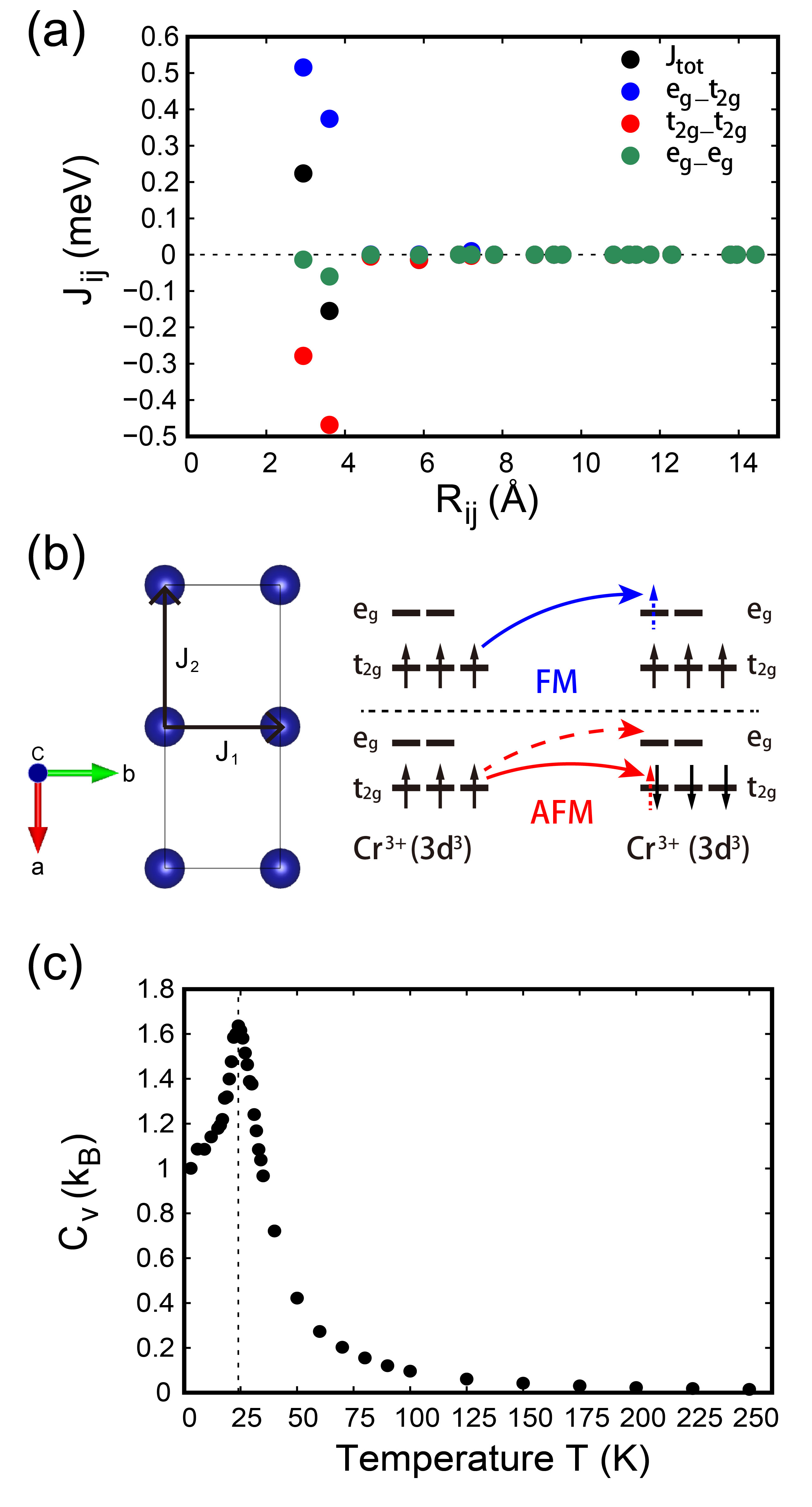}
\caption{(a) Calculated orbital-decomposed exchange parameters as a function of distance for Cr-Cr pairs in CrF$_{3}$. Positive and negative values correspond to FM and AFM coupling, respectively. (b) Schematic picture of interatomic hopping between Cr-Cr pairs. (c) The normalized specific heat versus temperature, calculated using the classical Monte Carlo method. The black dashed vertical line indicates T$_N$.}
\label{figure3}
\end{figure}

A mechanically stable 2D material should satisfy the Born-Huang criteria: the elastic modulus tensor components C$_{11}$, $C_{22}$, and $C_{66}$ should be positive, and $|C_{11}+C_{22}| > |2C_{12}$|. We fitted the curves of the energy changes U versus strains $\tau_{x y}$ using formula \cite{PhysRevB.85.125428,Zhang201416591}:
\begin{equation}
U=\frac{1}{2} C_{11} \tau_{x}^{2}+\frac{1}{2} C_{22} \tau_{y}^{2}+C_{12} \tau_{x} \tau_{y}+2 C_{66} \tau_{x y}^{2}.
\label{eq2}
\end{equation}
The components of stiffness tensors are C11=50.603 N/m, C22=100.686 N/m, C12=2.611 N/m, and C66=11.882 N/m. The results meet the criteria, confirming that r-CrF$_{3}$ monolayer are mechanically stable.



As shown in Fig. \ref{figure3}(a), the calculated magnetic exchange parameters J$_{ij}$, including orbital-decomposed contributions, are plotted as a function of the pairwise Cr-Cr distance. 
The first observation is that the exchange values (J$_{ij}$) are dominated by the first two nearest neighboring Cr-Cr pairs,
while the others ($R_{ij}> 4$ \AA) decrease rapidly to zero. 
This behavior exemplifies the pronounced localized nature of 3$d$ electrons on the Cr atoms, further supporting the validity of the Heisenberg model adopted in this study.
As the CrF$_3$ has a rectangular unitcell with different Cr-Cr lattice distances along ${\bf S}_y$ and ${\bf S}_x$, being 2.94 {\AA} and 3.61 {\AA}, the corresponding exchange parameters, as shown in Fig.~\ref{figure3}, are $J_1$ and $J_2$, respectively.
The first nearest-neighbor exchange coupling $J_1$ exhibits FM behavior with a value of 0.658 meV, whereas the second nearest-neighbor coupling $J_2$, despite its longer Cr-Cr distance, manifests AFM behavior with a robust magnitude of 1.986 meV.
This counterintuitive result can be effectively explained by the orbital-decomposed $J_{ij}$ data and the corresponding hopping mechanism.
The overall exchange coupling on Cr-Cr pair is dictated by the competition happens on different hopping forms.
For the Cr$^{3+}$ state, with three d-electrons remaining on each cation site, $t_{2g}$-$e_g$ hopping is the only allowed mechanism when the coupling between two Cr atoms is FM (upper panel in Fig. 3(b)), instead of $t_{2g}-t_{2g}$, which is restricted due to the Pauli exclusion principle. In contrast, for AFM coupling (lower panel), although both $t_{2g}-t_{2g}$ and $t_{2g}-e_g$ hopping paths are feasible, the $t_{2g}-t_{2g}$ is more energetically favorable due to the first Hund's rule.
This is perfectly consistent with the calculated data shown in Fig.~\ref{figure3}(a). 
For the first nearest-neighbor coupling $J_{1,\rm total}$ (black circles), its weak FM value of 0.658 meV arises from the competition between a positive $J_{1,e_g-t_{2g}}$ of 2.590 meV (green circles) and a negative $J_{1,t_{2g}-t_{2g}}$ of -1.895 meV (red circles).
In contrast, for the second nearest-neighbor coupling, due to a $31\%$ weaker $J_{2,e_g-t_{2g}}$ (reaching 1.788 meV) and a dominant $90\%$ increase in $J_{2,t_{2g}-t_{2g}}$ (reaching -3.597),
the system exhibits a net enhancement of AFM coupling $J_2$. 

\begin{figure}[htbp]
\includegraphics[scale=1]{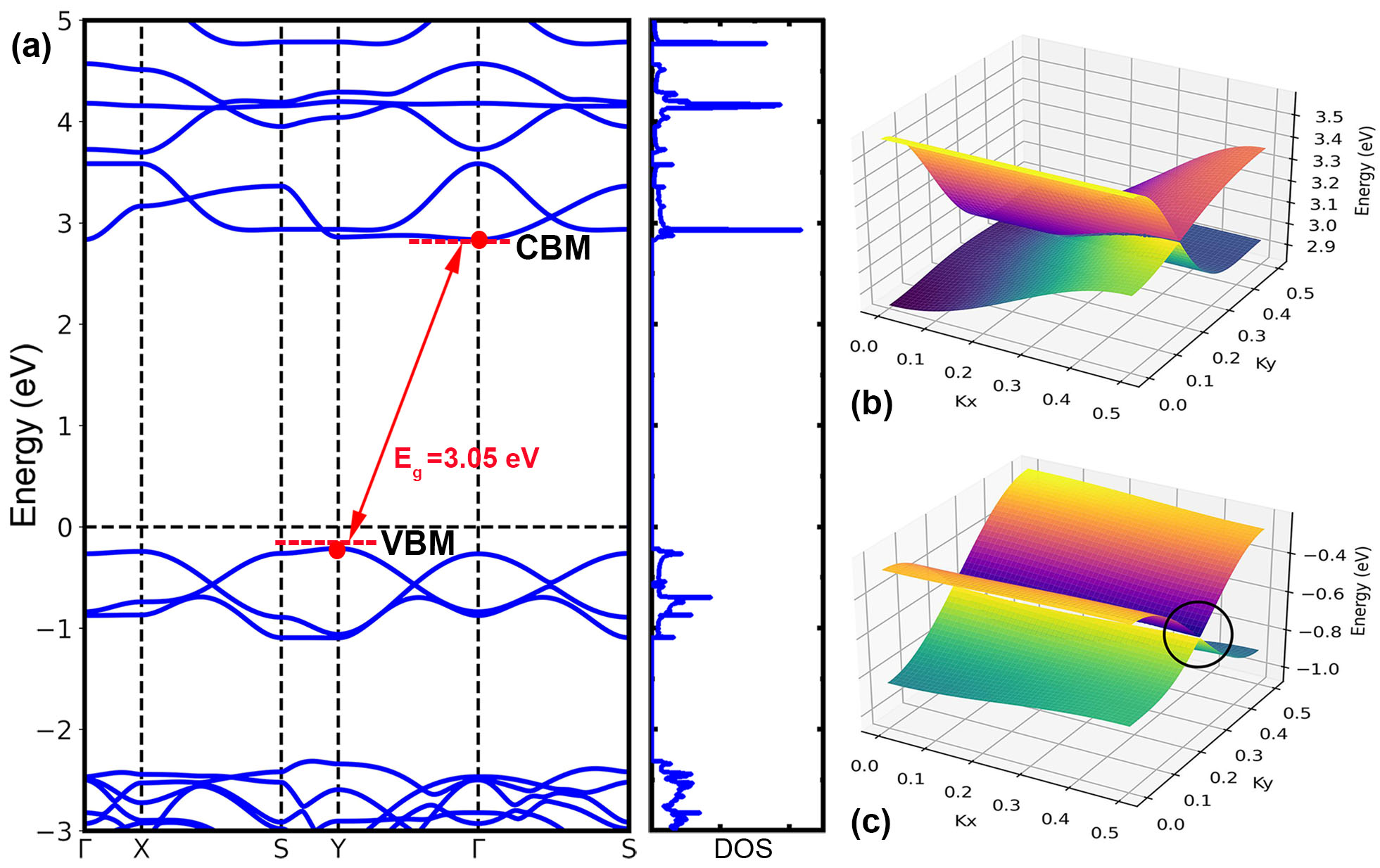}
\caption{(a) The band structure and total density of states of the r-CrF$_{3}$ monolayer. The spin-up and spin-down channels are degenerate. The VBM and CBM are indicated by red points. (b) The two conduction bands near the Fermi level in the first BZ. (c) The two valence bands near the Fermi level in the first BZ.}
\label{figure4}
\end{figure}

Using the calculated exchange parameters as input, we further determined the critical temperature of CrF$_3$ through both mean field approximation and classical Monte Carlo simulations.
The calculated $T_N$ are 48.3 K and 24 K, with the latter represented by the specific heat as a function of temperature, shown in Fig.~\ref{figure3}(c).
This result is comparable to the experimental and theoretical findings for monolayer CrI$_3$, which has a $T_c$ of 45 K~\cite{Huang2017,10.1021/acs.jpcc.1c04311}.


The electronic band structure and density of states (DOS) of the antiferromagnetic 2D r-CrF$_{3}$ are shown in Fig. \ref{figure4}. As illustrated in Fig. \ref{figure4} (a), the band structure reveals that CrF$_{3}$ is an indirect-gap semiconductor with a bandgap of 3.05 eV. The valence band maximum (VBM) is located at the Y point, while the conduction band minimum (CBM) is located at the $\Gamma$ point in the Brillouin zone. The corresponding DOS highlights the significant electronic states near the CBM, further confirming the electronic localization in this material. The three-dimensional band dispersions in Fig. \ref{figure4} (b) and (c) provide additional insights into the anisotropic electronic properties. Fig. \ref{figure4} (b) shows the conduction band near the CBM, exhibiting a nearly-flat shape, while Fig. \ref{figure4} (c) reveals a highly anisotropic saddle-like feature near the VBM (circled region), indicative of strong anisotropy in the valence band. Notably, the two valence bands near VBM cross at (0, 0.25) and (0.5, 0.25), forming two highly anisotropic Dirac points. This feature is reminiscent of the electronic behavior near the Fermi surface of rectangular carbon nitride (C$_{4}$N) monolayers \cite{adfm.202415606}, where quasi-one-dimensional dispersive bands dominate the transport properties. This anisotropy aligns with the quasi-one-dimensional nature of r-CrF$_{3}$'s mechanical and magnetic properties. Such a unique electronic structure, coupled with the sizable indirect bandgap and antiferromagnetic ordering, makes r-CrF$_{3}$ a promising material for applications in spintronic and low-dimensional electronic devices.


\begin{figure}[ht!]
    \centering
     \includegraphics[scale=0.58]{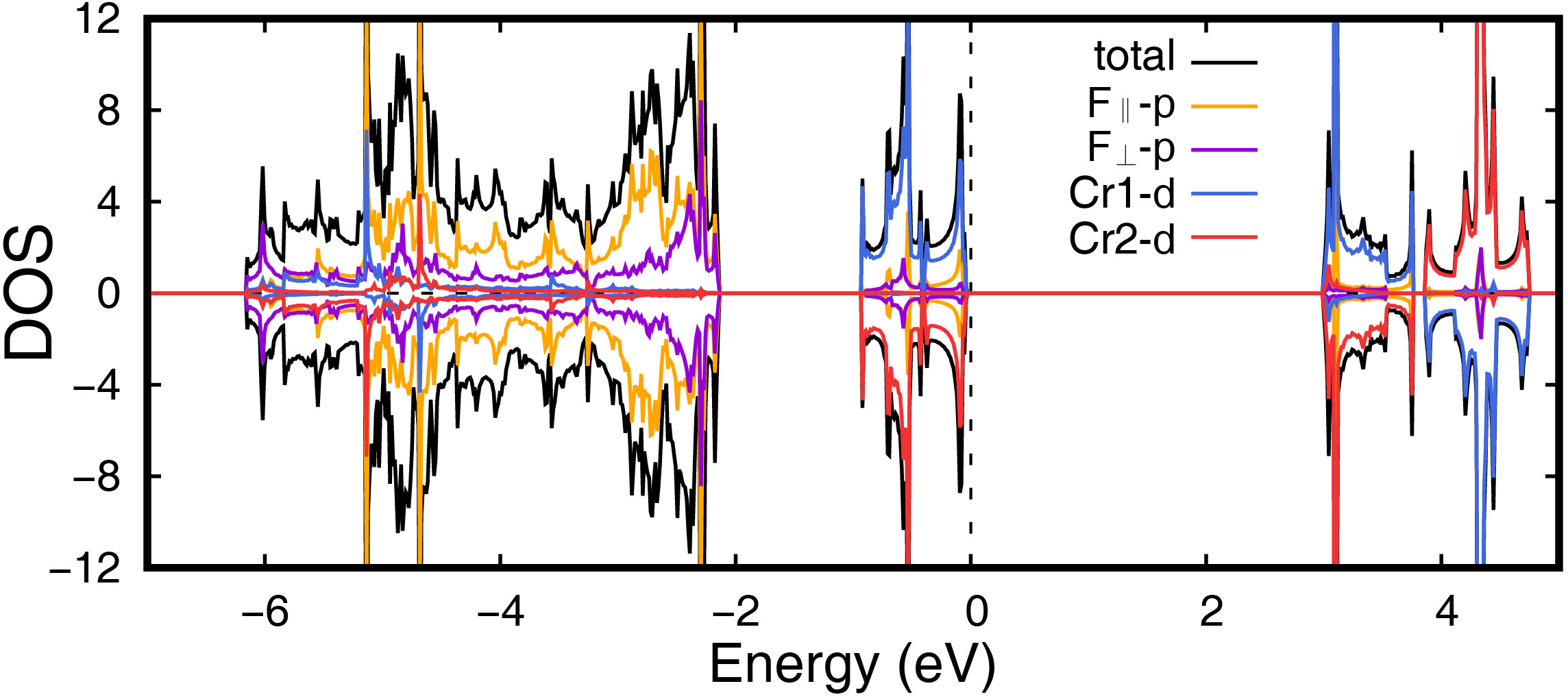}
			\caption{
     The calculated total DOS and the atomically decomposed DOS of CrF$_3$.}
    \label{figure5}
\end{figure}

The calculated total and atomically decomposed DOS are shown in Figure~\ref{figure5}. The most occupied Cr-$d$ states are distributed just below the Fermi level, within the energy range of -1 to 0 eV. The two Cr atoms in the unit cell exhibit opposite majority spin channel occupation due to their AFM nature. In addition, some minor induced states appear in the energy range of -6 to -3 eV as a result of hybridization between Cr and F atoms.

\begin{figure}[h]
    \centering
     \includegraphics[scale=0.65]{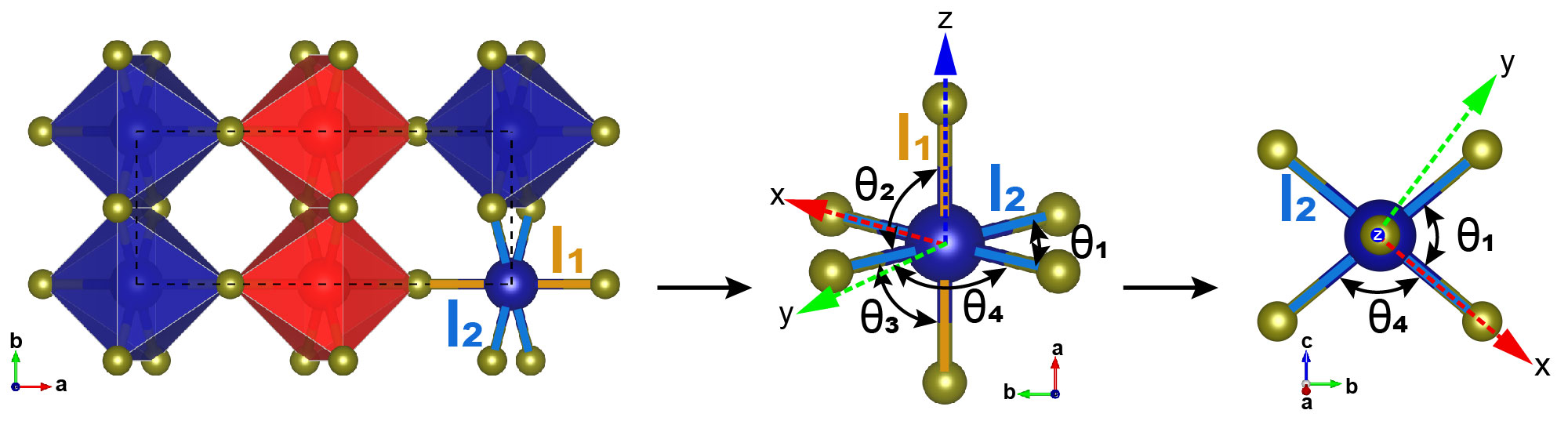}
			\caption{
     The top view of the r-CrF$_{3}$ monolayer structure (left panel), the local CrF$_{6}$ octahedron (middle panel), and the distorted $xy$-plane (right panel) . Blue and red polyhedral represent Cr1F$_{6}$ and Cr2F$_{6}$ octahedron, where the Cr1 and Cr2 construct the antiparallel spin chain along \textbf{a} direction with up spin in Cr1 and down spin in Cr2 atom. l and $\theta$ denote different bond lengths and bond angles. Dashed red, green, and green lines represent local basis in the octahedron.}
    \label{figure6}
\end{figure}

\begin{table}[!ht]

\caption{\label{tab:table2}{
Structural information of CrF$_{6}$ for the magnetic ground state.
}}
    \centering
 \begin{tabular}{|c|c|c|c|c|c|c|}\hline
 \hline
\textbf{}  & \multicolumn{4}{c|}{bond angle ($^{\circ}$)} & \multicolumn{2}{c|}{bond length (Å)}\\ \hline
\textbf{} &  $\theta$$_{1}$ & $\theta$$_{2}$   & $\theta$$_{3}$ & $\theta$$_{4}$   & l$_{1}$  & l$_{2}$ \\ \hline
Cr1  & 79.7   & 89.6   & 90.4  & 100.3  & 1.90          & 1.91          \\ \hline
Cr2  & 79.7   & 89.6   & 90.4   & 100.3 & 1.90          & 1.91          \\ \hline
\hline
\end{tabular}
\end{table}

As shown in the middle panel of Figure~\ref{figure6}, each Cr atom is surrounded by six ligand F atoms, forming an identical CrF$_6$ octahedral environment. 
In each of them, the six Cr-F bonds can be classified into two types based on their lengths (data shown in Table~\ref{tab:table2}): two shorter bonds ($l_1 \approx 1.90$ {\AA}) along $a$ direction and four longer bonds ($l_2 \approx 1.91$ {\AA}) in the $bc$ plane, representing a slight tetragonal compression strain with main axis along $a$ direction. 
The out-of-plane bond angles ($\theta_2$ and $\theta_3$) are nearly 90$^\circ$, with only a 0.4$^\circ$ deviation. In contrast, the angles in the $bc$ plane exhibit noticeable deviation, with two of them measuring 79.7$^\circ$ ($\theta_1$) and the other two 100.3$^\circ$ ($\theta_4$), indicating further symmetry breaking. Consequently, the originally square-shaped $bc$ plane transforms into a rectangular one, which is illustrated in the right panel in Figure~\ref{figure6}.
This low-symmetry, distorted octahedron structure is expected to induce a non-standard crystal field splitting similar to that of an ideal octahedral environment, as confirmed by the following DOS analysis.

\begin{figure*}[ht!]
    \centering
     \includegraphics[scale=1]{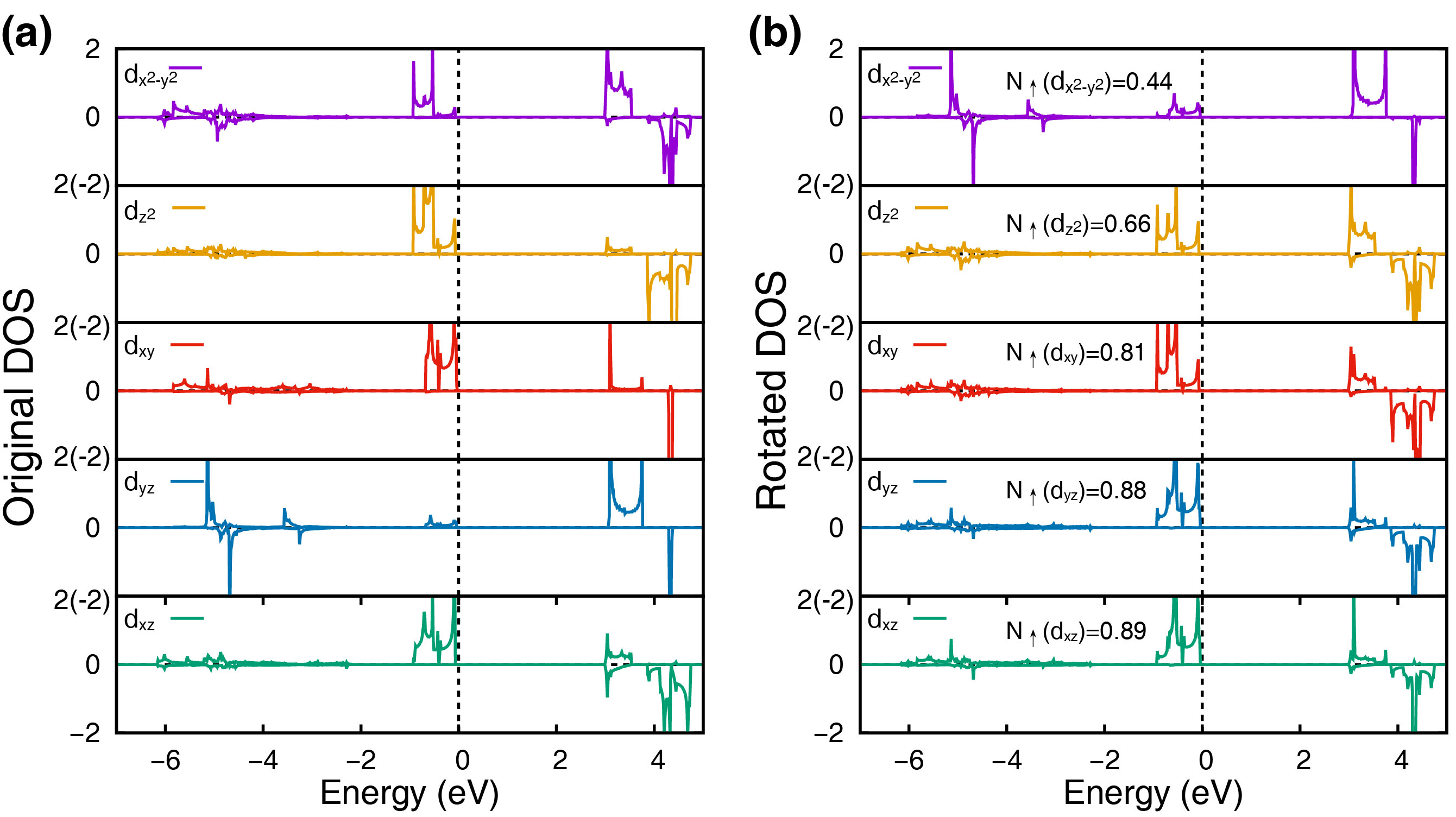}
			\caption{Orbital-decomposed DOS of Cr1 projected in (a) the global basis and (b) the local CrF$_6$ basis. The data for Cr2 are exactly the same, but with opposition spin polarization direction. Numbers shown in the figure represent the integrated DOS up to the Fermi level in the spin up channel.}
    \label{figure7}
\end{figure*}

\begin{table}[!ht]
\centering
\caption{The integrated DOS up to the Fermi level for the five decomposed $d$-orbits and the summed values of t$_{2g}$ and e$_{g}$ of $d$ states in the newly rotated CrF$_{6}$ basis. The magnetic moment, M, is calculated as N$_\uparrow$-N$_\downarrow$.}
\label{tab:table3}
\begin{tabular}{|cc|r|r|r|r|r|r|r|r|}
\hline\hline
 & & \(d_{xy}\) & \(d_{yz}\) & \(d_{xz}\) & \(d_{z^2}\) & \(d_{x^2-y^2}\) & \(t_{2g}\) & \(e_{g}\) & \(d_{total}\) \\
\hline
 & \(N_{\uparrow}\)
   & 0.81 & 0.88 & 0.89 & 0.66 & 0.44 & 2.57 & 1.10 & 3.67 \\
\text{Cr1} & \(N_{\downarrow}\)
   & 0.11 & 0.09 & 0.10 & 0.14 & 0.20 & 0.30 & 0.34 & 0.64 \\
 & \(M\)
   & 0.70 & 0.78 & 0.79 & 0.52 & 0.24 & 2.27 & 0.76 & 3.03 \\
\hline
 & \(N_{\uparrow}\)
   & 0.11 & 0.09 & 0.10 & 0.14 & 0.20 & 0.30 & 0.34 & 0.64 \\
\text{Cr2} & \(N_{\downarrow}\)
   & 0.81 & 0.88 & 0.89 & 0.66 & 0.44 & 2.57 & 1.10 & 3.67 \\
 & \(M\)
   & -0.70 & -0.78 & -0.79 & -0.52 & -0.24 & -2.27 & -0.76 & -3.03 \\
\hline\hline
\end{tabular}
\end{table}

In order to properly understand the $lm$-decomposed Cr DOS, we rotated the DOS data from the original global basis to a CrF$_6$ local basis. 
As shown in Figure~\ref{figure6}, the new basis are defined such that the $z$-direction aligns with the $a$-axis, coinciding with the shorter Cr-F bond, while $x$ and $y$ lie in the $bc$-plane. Due to the rectangular distortion within the plane, the $x$-direction is aligned along a specific Cr-F bond, whereas the $y$-direction deviates by approximately $10^\circ$ from the other Cr-F bond.
The results are shown in Figure~\ref{figure7}. In the original basis (panel (a)), the five $d$-states lack a well-defined physical interpretation. As a result, the DOS distribution and respective intensity vary randomly among the five $d$-orbitals, making it impossible to draw meaningful conclusions.
In contrast, significant alterations appear when the data are projected onto a properly chosen local basis.
First, the DOS distributions of the five $d$-orbitals can be classified into two categories: one including $d_{xy}$, $d_{yz}$, $d_{xz}$, and $d_{z^2}$ (the lower four panels), which exhibit similar spectra, whereas the other, $d_{x^2-y^2}$ (the top panel), is distinct from the rest.
This phenomenon is directly related to the in-plane distortion, 
which perturbs the system and induces a distinct Cr-F hybridization due to the misalignment between the $y$-direction and the bonding direction.
Second, among the four $d$-orbitals in the first category, the three $t_{2g}$ states exhibit a stronger DOS intensity below the Fermi level, indicating a higher degree of electron occupancy. This is further confirmed by the DOS integration results shown in Table~\ref{tab:table3}.
For the $d_{xy}$, $d_{yz}$, and $d_{xz}$ orbitals, the occupied electron numbers in spin-majority channel are 0.81, 0.88, and 0.89 $e$, respectively, which are higher than those of $d_{z^2}$ (0.66 $e$) and $d_{x^2-y^2}$ (0.44 $e$) orbitals. These values align well with the octahedral crystal field splitting, where the three-fold $t_{2g}$ states are nearly half-filled, with a total of 2.57 $e$, while the two-fold $e_g$ states contain significantly fewer electrons, totaling 1.10 $e$. The latter is referred to as induced moments, which can be traced back to the hybridization between Cr and ligand F atoms and is a common phenomenon, as it has been observed in other Cr-based compounds\cite{10.1038/s41699-024-00490-9,10.1021/acs.jpcc.1c04311}.

\begin{figure}[htbp]
\includegraphics[scale=1]{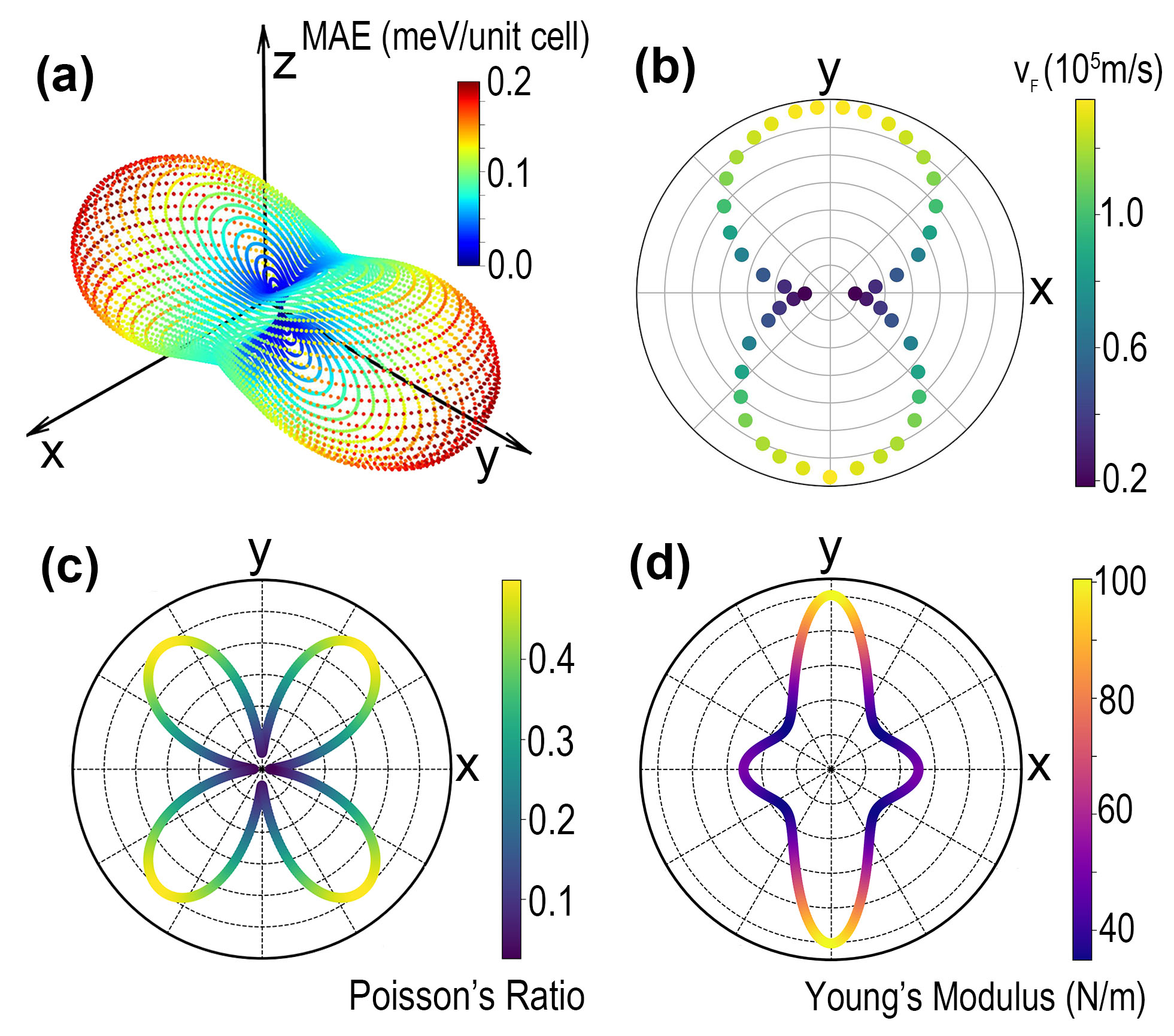}
\caption{(a) Directional dependence of the MAE in r-CrF$_3$. (b) Fermi velocity distribution of the type-II Dirac point in the valence bands. (c) Polar plot of the in-plane Poisson's ratio of r-CrF$_3$. (d) Polar plot of the in-plane Young's modulus of r-CrF$_3$.}
\label{figure8}
\end{figure}

The anisotropic properties of the r-CrF$_3$ monolayer are clearly revealed in its magnetic, electronic, and mechanical behaviors, as shown in Fig. \ref{figure8}. Panel (a) illustrates the MAE distribution, where larger values correspond to energetically unfavorable magnetic orientations. MAE is defined as the difference between energies corresponding to the magnetization along the sepecific direction and along the easy-axis ($MAE=E_{\vec{r}} - E_{easy}$). The MAE reaches its minimum along the out-of-plane (z-axis) direction, indicating that r-CrF$_3$ prefers an out-of-plane magnetic structure, which is a key feature of its antiferromagnetic ordering. The MAE reaches its maximum along the y-axis direction, of 0.098 meV/Cr atom. In contrast, the value is 0.025 meV/Cr atom in CrCl$_{3}$, 0.160 meV/Cr atom in CrBr$_{3}$, and 0.804 meV/Cr atom in h-CrI$_{3}$ \cite{PhysRevB.98.144411}. The large MAE in h-CrI$_{3}$ and h-CrBr$_{3}$ is attributed to the strong spin-orbit coupling (SOC) in the heavier iodine ions, while the greater MAE in r-CrF$_{3}$ compared to h-CrCl$_{3}$ arises from symmetry breaking and strong structural anisotropy. The  Fig. \ref{figure8} (b) presents the Fermi velocity distribution for the type-II Dirac point in the valence bands, showing significant anisotropy. The Fermi velocity is markedly higher along the y-direction compared to the x-direction, suggesting quasi-one-dimensional electronic transport properties near the valence band edge. 

The direction-dependent in-plane Young's moduli and Poisson's ratios are computed using the two formulae \cite{PhysRevB.85.245434,PhysRevB.82.235414}:
\begin{widetext}
\begin{equation}
E_{2D}(\theta)=\frac{C_{11} C_{22}-C_{12}^{2}}{C_{11} \sin ^{4} \theta+C_{22} \cos ^{4} \theta+\left(\frac{C_{11} C_{22}-C_{12}^{2}}{C_{66}}-2 C_{12}\right) \cos ^{2} \theta \sin ^{2} \theta}, 
\end{equation}
and
\begin{equation}
\vartheta(\theta)=-\frac{\left(C_{11}+C_{22}-\frac{C_{11} C_{22}-C_{12}^{2}}{C_{66}}\right) \cos ^{2} \theta \sin ^{2} \theta-C_{12}\left(\cos ^{4} \theta+\sin ^{4} \theta\right)}{C_{11} \sin ^{4} \theta+C_{22} \cos ^{4} \theta+\left(\frac{C_{11} C_{22}-C_{12}^{2}}{C_{66}}-2 C_{12}\right) \cos ^{2} \theta \sin ^{2} \theta},
\end{equation}
\end{widetext}
in which $\theta$ is the angle of the arbitrary direction and x-direction, $E_{2D}(\theta)$ and $\vartheta(\theta)$ are in-plane Young's moduli and Poisson's ratios in $\theta-$direction. Fig. \ref{figure8} (c) reveals the in-plane Poisson's ratio, which is nearly zero along both the x- and y-directions, implying that the r-CrF$_3$ monolayer exhibits zero Poisson's ratio in-plane. This unusual mechanical property suggests that when a uniaxial strain is applied along one axis, the material undergoes minimal lateral deformation, making it highly stable and desirable for mechanical applications requiring robust in-plane responses. Finally, Fig. \ref{figure8} (d) displays the in-plane Young's modulus, which is highly anisotropic, with significantly larger stiffness along the y-direction compared to the x-direction, further confirming the structural anisotropy arising from the rectangular lattice. These combined results highlight the strongly anisotropic nature of r-CrF$_3$ in its magnetic, electronic, and mechanical properties, with its out-of-plane magnetic preference and zero in-plane Poisson's ratio making it a promising material for advanced spintronic and mechanical applications.

\begin{figure}[htbp]
\includegraphics[scale=1]{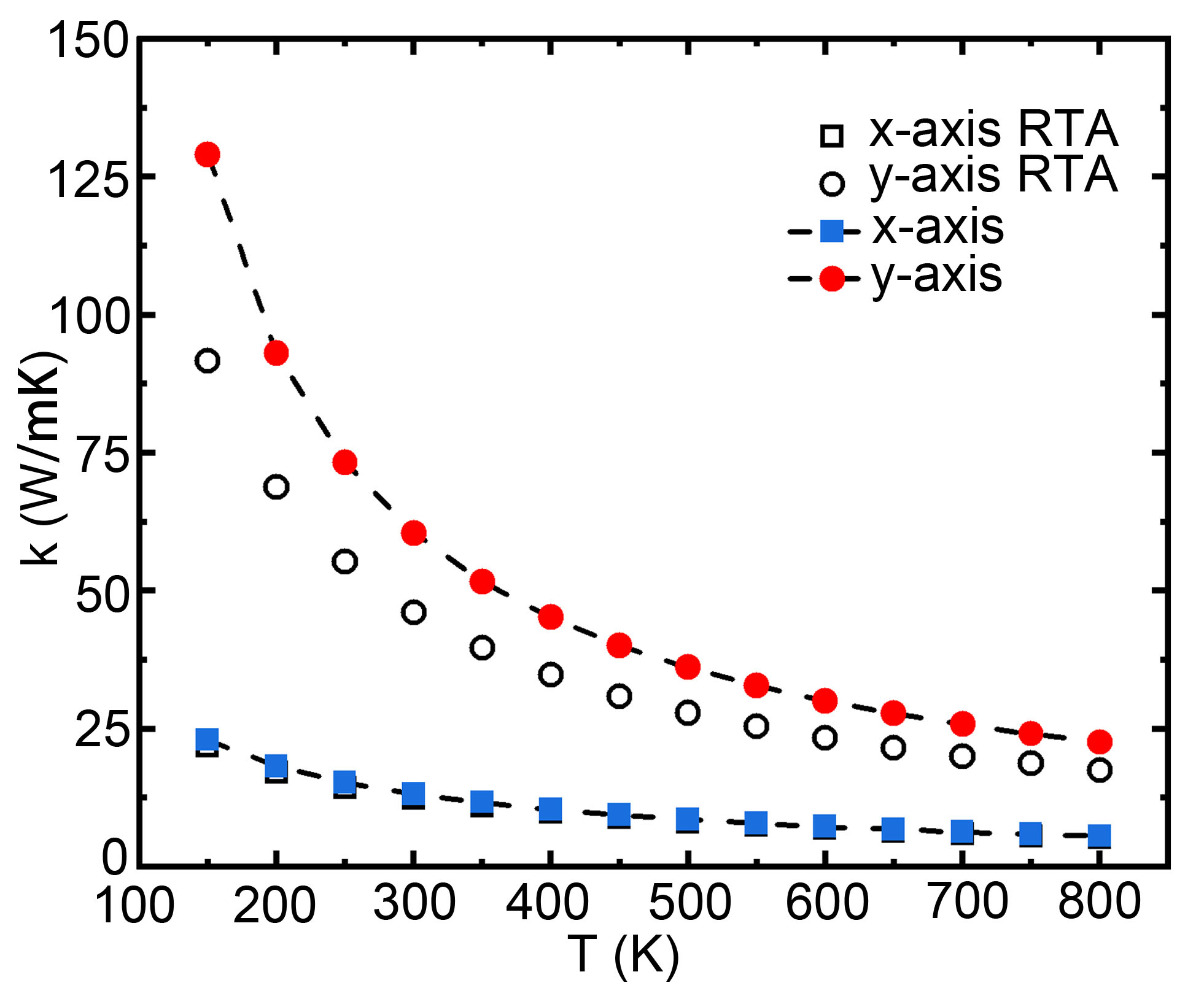}
\caption{The thermal conductivities of monolayer r-CrF$_{3}$ along x-axis and y-axis direcitons calculated with RTA and from the iterative solution of BTE.
}
\label{figure9}
\end{figure}

The thermal conductivity $\kappa$ as a function of temperature of monolayer r-CrF$_{3}$ are shown in Fig. \ref{figure9}. It demonstrates pronounced anisotropic heat transport, with significantly higher values along the \(y\)-axis compared to the \(x\)-axis across all temperatures. The relaxation time approximation (RTA) results and iterative BTE results are nearly identical along the \(x\)-axis. In contrast, along the \(y\)-axis, the iterative BTE solution yields higher \(k\) values than the RTA. At \(300 \, \text{K}\), \(\kappa_{y}\) is approximately \(60.5 \, \text{W/mK}\), while \(\kappa_{x}\) is only \(13.2 \, \text{W/mK}\). The thermal anisotropic factor is defined as $f_{iso}=\kappa_{max}/\kappa_{min}$, and the factor of r-CrF$_{3}$ is computed to be 4.58. By contrast, $f_{iso}$ of black phosphorene and WTe$_{2}$ are of \(2 \sim 3\) \cite{Ma_2016,PhysRevB.90.214302}, that of arsenene is about 4 \cite{PhysRevB.93.085424,10.1063/5.0021237}. This emphasizes the highly anisotropic nature of \(r\)-CrF$_{3}$ and its potential for directional heat management applications.

\begin{figure}[htbp]
\includegraphics[scale=1]{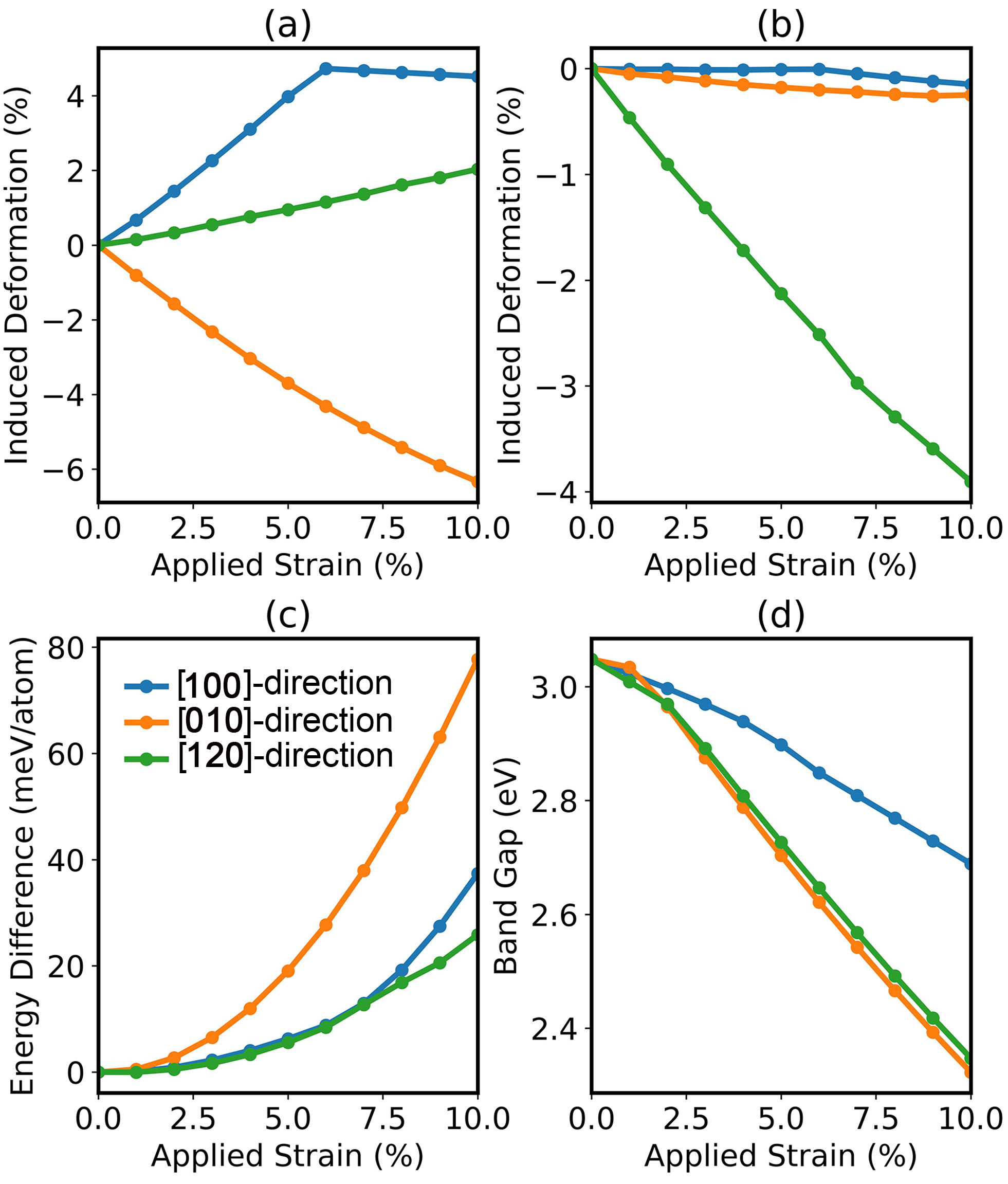}
\caption{(a) In-plane strain-induced deformation along the out-of-plane (\textit{z}) direction. (b) In-plane strain-induced deformation along the perpendicular in-plane direction. (c) Energy change as a function of applied in-plane strain. (d) Band gap variation under in-plane strain. The blue, orange, and green lines correspond to the [001], [010], and [120] directions, respectively.
}
\label{figure10}
\end{figure}

The mechanical and electronic responses of r-CrF$_3$ under in-plane strain exhibit unique and intriguing properties, as shown in Fig. \ref{figure10}. In panel (a), the strain-induced deformation along the out-of-plane direction reveals highly anisotropic Poisson’s ratio behavior. When tensile strain is applied in the y-direction ([010] direction), the thickness of the monolayer decreases, exhibiting a positive Poisson's ratio. In contrast, when tensile strain is applied in the x-direction ([100] direction) or the [120]] direction, the thickness increases, resulting in a negative Poisson's ratio. This NPR is a rare phenomenon where the material expands in the out-of-plane direction under in-plane tension, a behavior characteristic of auxetic materials. Such materials are highly valued for their superior toughness, energy absorption, and resistance to fracture, making r-CrF$_3$ an excellent candidate for mechanical metamaterials and flexible devices.

Fig. \ref{figure10} (b) highlights the strain-induced deformation in the perpendicular in-plane direction and demonstrates a remarkable feature of r-CrF$_3$: minimal coupling between orthogonal in-plane directions. When tensile strain is applied along the x- or y-direction, the deformation in the perpendicular direction is negligibly small, indicating that the response to strain is largely confined to the strain direction itself. This quasi-one-dimensional mechanical nature suggests that r-CrF$_3$ behaves as a highly decoupled mechanical system in-plane, where strain in one direction does not induce significant deformation in the perpendicular direction. Such behavior is rare in conventional 2D materials and points to unique lattice dynamics in r-CrF$_3$ that could enable applications requiring directional mechanical responses or strain isolation.

\begin{figure*}[htbp]
\includegraphics[scale=1]{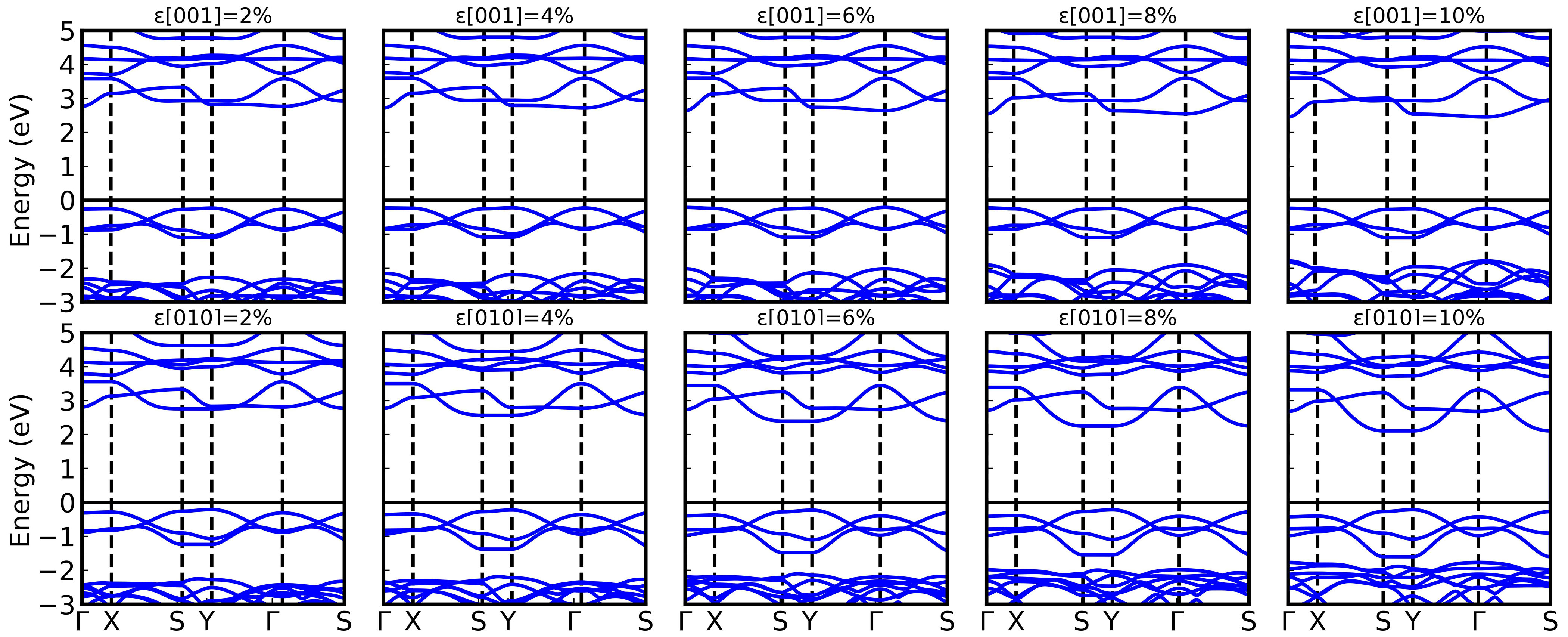}
\caption{Energy band structures of r-CrF$_3$ under uniaxial tensile strain. The strain levels are 2\%, 4\%, 6\%, 8\%, and 10\% from left to right. The upper row corresponds to strain applied in the x-direction ([001]), while the lower row corresponds to strain applied in the y-direction ([010]).}
\label{figure11}
\end{figure*}

The strain-induced energy variation is shown in Fig. \ref{figure10} (c), which increases smoothly and continuously with strain, with no evidence of phase transitions. This mechanical stability under strain ensures that r-CrF$_3$ retains its structural integrity over a broad range of applied strains, an essential property for practical strain-engineering applications. 

As shown in Fig.~\ref{figure10}(d), the band gap decreases monotonically as the tensile strain increases in each of the three in-plane directions considered. Notably, the gap reduction is less pronounced when strain is applied along the $x$-axis, whereas strain along the $y$- and [120] directions induces a more significant decrease. This observation reflects how mechanical deformation along each crystallographic direction differently affects the conduction-band features near the Fermi level. In particular, strain along [120] lowers both the conduction-band minimum (CBM) and the next conduction band (CBM--1) similarly to the behavior under $y$-axis strain, resulting in a comparably large band-gap reduction. By contrast, $x$-direction strain impacts primarily the CBM, thereby yielding a relatively smaller gap decrease. This directional dependence underscores that the electronic properties of r-CrF$_3$ can be sensitively tuned through strain, especially when applied along specific in-plane axes.

The energy band evolution of r-CrF$_3$ under uniaxial tensile strain, as shown in Fig. \ref{figure11}, reveals a distinct and independent response of the two lowest-energy conduction bands near the Fermi level. These two conduction bands can be distinguished by their energies at the $\Gamma$ point: the lower-energy conduction band at $\Gamma$ is more sensitive to strain applied in the x-direction ([001]), whereas the higher-energy conduction band at $\Gamma$ primarily responds to strain in the y-direction ([010]). When uniaxial strain is applied along the x-direction, the lower conduction band shifts downward significantly, while the higher-energy band remains almost unchanged. Conversely, under strain in the y-direction, the higher-energy conduction band shifts downward, with minimal impact on the lower-energy conduction band. This behavior indicates a decoupling of the two conduction bands, where each band is selectively regulated by strain along a specific in-plane direction, with negligible influence from strain in the orthogonal direction.

This directional and independent regulation of the conduction bands can be linked to the earlier PDOS analysis and the minimal coupling between orthogonal directions. The sensitivity of the lower-energy band to x-direction strain likely arises from the interaction between the Cr-d$_{xz}$ orbital and the F orbitals, while the higher-energy band is regulated primarily by the coupling of the Cr-d$_{yz}$ orbital with the F-p$_y$ orbital under y-direction strain. The minimal coupling between these orthogonal directions allows the conduction bands to respond independently to uniaxial strain, reinforcing the quasi-one-dimensional nature of the electronic structure.

The independent strain response of the two lowest-energy conduction bands near the $\Gamma$ point is a special feature of r-CrF$_3$, offering significant potential for directional strain engineering of its electronic properties. By selectively tuning strain along the x- or y-direction, it becomes possible to manipulate the conduction states without affecting those governed by the orthogonal direction. This decoupled, anisotropic response positions r-CrF$_3$ as a promising material for strain-tunable electronic and optoelectronic devices, where independent control of conduction channels could enable novel functionalities.

\section{Summary}
Using first-principles calculations combined with an evolutionary structure search, we report the discovery of a rectangular phase of CrF$_3$, namely r-CrF$_3$, that is more stable than the widely assumed hexagonal phase (h-CrF$_3$). This finding challenges the conventional understanding of CrF$_3$, which has long been believed to adopt the hexagonal structure seen in h-CrI$_3$, h-CrBr$_3$, and h-CrCl$_3$. The identification of the stable rectangular phase introduces a new perspective on the structural and physical properties of CrF$_3$. The rectangular phase of CrF$_3$ exhibits remarkable anisotropic features. Mechanically, r-CrF$_3$ shows quasi-1D behavior with nearly zero Poisson’s ratio under uniaxial strain along the x- and y-directions, with minimal coupling between orthogonal directions. Electronically, its conduction bands near the Fermi level are decoupled, allowing independent strain engineering along the x- and y-directions. The bandgap narrows under tensile strain, enhancing its potential for strain-engineered electronics. R-CrF$_3$ also exhibits a rare negative Poisson’s ratio under strain, making it a promising candidate for mechanical metamaterials. Magnetically, it maintains out-of-plane antiferromagnetic ordering with a magnetic anisotropy energy of 0.098 meV/Cr atom and an estimated Néel temperature of 24 K. Thermally, r-CrF$_3$ displays significant thermal anisotropy, with thermal conductivity of 60.5 W/mK along the y-direction at 300 K, compared to 13.2 W/mK along the x-direction. This highly directional heat transport, combined with its quasi-1D mechanical and electronic properties, suggests r-CrF$_3$ is suitable for thermal management applications. In conclusion, r-CrF$_3$ combines structural stability, quasi-1D mechanical and electronic properties, NPR, strain-tunable electronic states, anisotropic thermal conductivity, making it a promising candidate for applications in strain-engineered electronics, mechanical metamaterials, and thermal management.

\begin{acknowledgments}
We would like to thank the Shandong Institute of Advanced Technology for providing computational resources. This work is supported by the National Natural Science Foundation of Shandong Province(Grant No. ZR2024QA040). Xin Chen thanks China scholarship council for financial support (No. 201606220031). Cheng Shao acknowledge support from  the National Natural Science Foundation of China (Grant No. 52306102), the Excellent Young Scientists Fund (Overseas) of Shandong Province (2023HWYQ-107), and the Taishan Scholars Program of Shandong Province. Duo Wang acknowledges financial support from the Science and Technology Development Fund from Macau SAR (Grant No. 0062/2023/ITP2) and the Macao Polytechnic University (Grant No. RP/FCA-03/2023). Biplab Sanyal acknowledges financial support from Swedish Research Council (grant no. 2022-04309), STINT Mobility Grant for
Internationalization (grant no. MG2022-9386) and DST-SPARC, India (Ref. No. SPARC/2019-2020/P1879/SL).
\end{acknowledgments}

\bibliography{Ref}

\end{document}